\pdfoutput=1
\documentclass[letterpaper, 10 pt, conference]{ieeeconf}  

\IEEEoverridecommandlockouts                              

\usepackage{caption}
\usepackage{cite}
\usepackage{amsmath,amssymb,amsfonts}
\usepackage{subfig}
\usepackage{graphicx}
\usepackage{textcomp}
\usepackage{xcolor}
\usepackage{booktabs}
\usepackage{multirow}
\usepackage[algo2e]{algorithm2e}
\usepackage{algorithm}  
\usepackage{algpseudocode}  
\usepackage{float} 

\def\BibTeX{{\rm B\kern-.05em{\sc i\kern-.025em b}\kern-.08em
    T\kern-.1667em\lower.7ex\hbox{E}\kern-.125emX}}

\title{\LARGE \bf
Decentralized Input and State Estimation for Multi-agent System with Dynamic Topology and Heterogeneous Sensor Network}

\author{Zida Wu$^{1}$,  Ankur Mehta$^{1}$ 
\thanks{ $^{1}$Department of Electrical and Computer Engineering, University of California Los Angeles, Los Angeles, CA, USA.
        {\tt\small \{zdwu,mehtank\} @ucla.edu}        }%
}

\begin{document}

\maketitle
\thispagestyle{empty}
\pagestyle{empty}

\begin{abstract}
A crucial challenge in decentralized systems is state estimation in the presence of unknown inputs, particularly within heterogeneous sensor networks with dynamic topologies. While numerous consensus algorithms have been introduced, they often require extensive information exchange or multiple communication iterations to ensure estimation accuracy. This paper proposes an efficient algorithm that achieves an unbiased and optimal solution comparable to filters with full information about other agents. This is accomplished through the use of information filter decomposition and the fusion of inputs via covariance intersection. Our method requires only a single communication iteration for exchanging individual estimates between agents, instead of multiple rounds of information exchange, thus preserving agents' privacy by avoiding the sharing of explicit observations and system equations. Furthermore, to address the challenges posed by dynamic communication topologies, we propose two practical strategies to handle issues arising from intermittent observations and incomplete state estimation, thereby enhancing the robustness and accuracy of the estimation process. Experiments and ablation studies conducted in both stationary and dynamic environments demonstrate the superiority of our algorithm over other baselines. Notably, it performs as well as, or even better than, algorithms that have a global view of all neighbors.
\end{abstract}

\section{INTRODUCTION}
\label{section1}

Centralized systems are prevalent in real-world applications, requiring detailed information synchronization across components. However, scenarios like large-scale spatial multi-agent systems or sensor networks face challenges with the reliability and availability of global information in dynamic settings. Decentralized systems, in comparison, demonstrate significant resilience against individual node failures, offer enhanced scalability, and reduce synchronization demands \cite{kia2019tutorial}. For clarity and consistency, this paper uses a 'decentralized' multi-agent system to refer to systems where agents dynamically access information only from their immediate neighbors.

Decentralized estimation focuses on developing a fusion algorithm that gathers data from adjacent sensors to achieve consensus among local nodes, often through a consensus algorithm. These algorithms are divided into consensus-based\cite{olfati2007consensus, olfati2008distributed, kamal2011generalized, kamal2012information}, gossip-based\cite{ma2016gossip}, and diffusion-based\cite{cattivelli2010diffusion, vahidpour2019partial} methods based on their interaction and fusion techniques, as categorized in recent literature \cite{he2020distributed}. In consensus-based algorithms, each node communicates with connected neighbors to compute a weighted average of received states, iterating until discrepancies are minimized. Gossip-based algorithms similarly collect messages iteratively but randomly and asynchronously from neighborhoods, avoiding data aggregation from all. Diffusion-based methods, distinct in requiring only a single iteration between observations, perform an average of states post one iteration. Those methods require strong requirements of full communication rate, large communication range and bandwidth, and bring in correlated problems. Some algorithms, like max consensus\cite{petitti2011consensus}, achieve consensus in one iteration by synchronizing states to a node within the neighborhood that owns the minimum variance; or based on pseudo-estimates \cite{govaers2012exact,govaers2015comparison} to achieve global optimal but requires strict full communication rate, large bandwidth, and lacks of local optimal.

\begin{figure}[htbp]        
 \center{\includegraphics[width=6cm]  {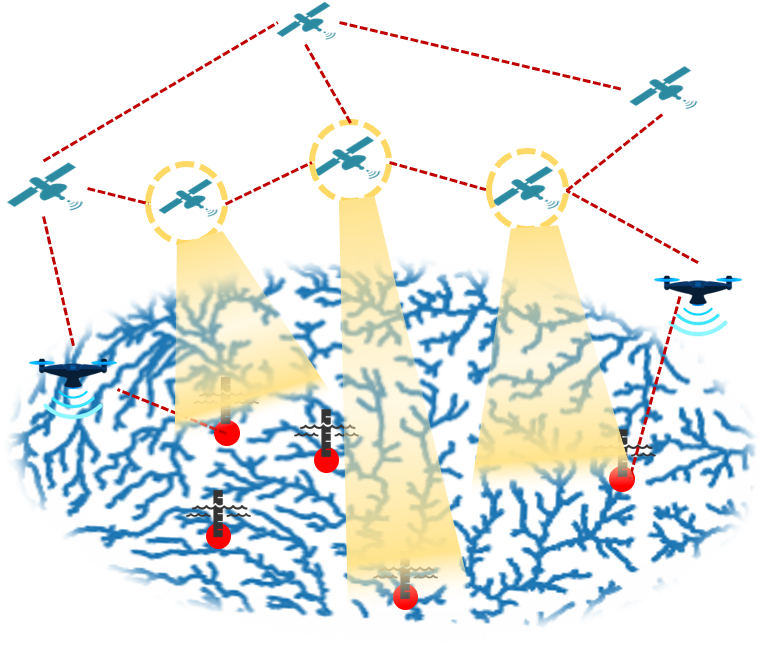}}     
 \caption{\label{demo} An example application of a decentralized estimation system is geophysical environment detection, which involves a multi-layered, heterogeneous sensor network with a dynamic topology. Gauges within the river reach offer localized measurements, albeit within a confined spatial range. Satellites afford broader spatial observations but lack real-time monitoring capabilities. Drones complement the observations by providing additional data points.}
\setlength{\belowcaptionskip}{-1cm}  
\vspace{-0.2cm} 
\end{figure}

Beyond state estimation, many scenarios require addressing unknown inputs affecting system dynamics, such as in geophysical detection, environmental monitoring\cite{ben2023multi}\cite{jaramillo2023decentralized}, maneuver tracking\cite{lee2015imm}, flight detection\cite{wu2022joint}, and sensor fault detection\cite{davoodi2013distributed}. These unknown inputs, often resulting from unforeseen events, need to be estimated alongside the system's state. The single-agent unknown input and state estimation has been well-explored in the recursive Kalman filter (RKF) \cite{gillijns2007unbiased}. Decentralized state estimation with unknown inputs generally adopts one of two strategies: input estimation using individual information \cite{peng2023optimal}, or assuming agents can access and share system equations and observations with neighbors \cite{liu2018minimum}, applying a local recursive Kalman filter for input estimation. However, the first approach lacks input consensus, while the second poses privacy risks due to the requirement of unrestricted bandwidth for data exchange and potential exposure of agent information.

This paper introduces a decentralized approach for input and state estimation within a sensor network featuring dynamic topology and heterogeneous sensors. Dynamic topology refers to the changing observation agents and neighborhoods over time, while heterogeneous sensors indicate different types of observations among agents. Instead of sharing system equations and observations, nodes only exchange their input and state estimations and covariances with neighbors. Input fusion utilizes covariance intersection \cite{he2018consistent, chang2021resilient, julier1997non, chen2002estimation}, and state fusion, through information filter decomposition \cite{durrant1990toward}, achieves performance comparable to full-information filters without the correlation issues that arise if all-to-all communication is allowed. Our method also includes specific operations, such as reviewing observation time windows and state compensation, to enhance estimation in dynamic topologies with a limited sensing range. Our algorithm requires only a one-time information exchange and does not need all-to-all communication or a full communication rate. The messages exchanged avoid exposing system matrices or raw measurements, thereby preserving privacy.

\section{Problem Formulation}
\label{section2}
\subsection{System description}\label{section:problem}
From the view of a single agent $i$, the system description is as follows:
\begin{equation}
{x_{k }^i} = {A_{k-1}^i}{x_{k-1}^i} + {G_{k-1}^i}{d_{k-1}^i} + {w_{k-1}^i} 
\label{eq1}
\end{equation}

\begin{equation}
{y_{k}^i} = {H_{k}^i}{x_{k}^i} + {v_{k}^i} 
\label{eq2}
\end{equation}
where the noise ${w_k^i}$ and ${v_k^i}$ are independent, zero-mean, white Gaussian noises with covariance matrices are $Q^i$ and $R^i$. We primarily focus on homologous states and inputs across agents, meaning system dynamics are consistent. Therefore for simplicity, we assume $A_k^i = A_k$, $G_k^i = G_k$, $Q_k^i = Q_k$, state $x_k^i\in {\mathbb{R}^n}$, input $d_k^i \in {\mathbb{R}^m}$, and output $y^i \in {\mathbb{R}^{p^i}}$ with varying $p^i$ across different observation models. In this system, both state $x_k^i$ and input $d_k^i$ are unknown, with each agent aiming to optimally estimate them.

Agents start with individual input estimation using the initial steps of the recursive Kalman filter (RKF) \cite{gillijns2007unbiased, wu2022joint}, where they predict the state and covariance according to Eq. (\ref{eq:info_kf_prediction0}) and estimate the input via Eq.(\ref{eq:input_estimation}).

\begin{equation}
\begin{aligned}
\hat{x}^i_{k \mid k-1} & =A^i_{k-1} \bar{x}^i_{k-1 \mid k-1}\\
\hat{P}^i_{k \mid k-1}
& =A^i_{k-1}  \bar{P}^i_{k-1 \mid k-1} A_{k-1}^{i^T}+Q^i_k
\end{aligned}
\label{eq:info_kf_prediction0} 
\end{equation}
Input estimation:
\begin{equation}
\hat{d}^i_{k-1}=M^i_{k}\left(y^i_{k}-H^i_{k} \hat{x}^i_{k \mid k-1}\right)
\label{eq:input_estimation}
\end{equation}
where $M^i_k$ is:
\begin{equation}
\begin{aligned}
   {M^i_k} = {[{({H^i_k}{G^{i}_{k - 1}})^T}{\tilde{R}_k^{ i^{- 1}}}({H^i_k}{G^i_{k - 1}})]^{ - 1}}{({H^i_k}{G^i_{k - 1}})^T}\tilde{R}_k^{ i^{- 1}} 
\end{aligned}
\label{eq:input_estimation_mk1}
\end{equation}
\begin{equation}
   \tilde{R}^i_k = H^i_k \hat{P}^i_{k \mid k-1} H^{i^T}_k+R^i_k
\label{eq:input_estimation_mk2}
\end{equation}

The covariance of input estimation is,
\begin{equation}
    \hat{S}^i_k(\hat{d}^i_k)={[{({H^i_k}{G^{i}_{k - 1}})^T}{( \tilde{R}^i_k)^{ - 1}}({H^i_k}{G^{i}_{k - 1}})]^{ - 1}}
\end{equation}

The rank condition for input estimation is the same as RKF,
\begin{equation}
    rank({H^i_k}{G^i_{k - 1}}) = rank({H^i_{k - 1}}) = m
\label{eq10}
\end{equation}
Following individual estimations, agents proceed differently from RKF by performing a pure Kalman filter (KF) update without including input estimation. This choice avoids extra computation as input estimation's inclusion does not alter the fusion result in the final algorithm but increases the computational load. The explanation will be detailed in the state fusion section.

\begin{equation}
\begin{aligned}
\tilde{x}^i_{k \mid k}
& =\hat{x}^i_{k \mid k-1}+L^i_{k}\left[y^i_{k}-H^i_{k}  \hat{x}^i_{k \mid k-1}\right] \\
\tilde{P}_{k \mid k}^{i^{-1}}
& =\hat{P}^{i^{-1}}_{k \mid k-1}+H^i_{k} R_{k}^{i^{-1}} H^i_{k} \\
L^i_{k}
& =\tilde{P}^i_{k \mid k} {H}^{i^{-1}}_{k}  {R}_{k}^{i^{-1}} \\
\end{aligned}
\label{eq:info_kf_update0}
\end{equation}

Each agent's estimation results in input and state estimations along with their covariances, presented as $\hat{d}^i_k$, $\hat{S}^i_k$, $\hat{x}^i_{k \mid k-1} $, $\hat{P}^i_{k \mid k-1}$, $\tilde{x}^i_{k \mid k}$, and $\tilde{P}_{k \mid k}^{i}$, adhering to the unbiased and minimum-variance (MVU) principle. Due to bandwidth and privacy limitations, transmitting direct observation or sharing agents' system dynamics or observation models is infeasible. This paper proposes a fusion algorithm using only these values to achieve an approximate optimal fusion result, bypassing the need for system information exchange. The fusion algorithm comprises two parts: input fusion and state fusion, which are explored in subsequent sections.

\section{Input fusion via Covariance Intersection}
Within the 1-hop distance of node $i$, including itself, represented by $\mathcal{N}_i$, communication is confined to $\mathcal{N}_i$. The correlation among received input estimations $d_j$ ($j \in \mathcal{N}^i$) is unknown, preventing optimal fusion akin to fully informed filters \cite{liu2018minimum}. We employ a method known as covariance intersection (CI) for fusion under uncertain correlation. CI, typically used in Kalman filter-based information fusion when source correlations are unclear, offers a way to merge information without complete correlation knowledge, albeit solving its non-linear optimization problems can be computationally intensive. This paper introduces a rapid CI approach \cite{he2018consistent, niehsen2002information, he2020distributed} to maintain fusion consistency and ensure real-time computational feasibility.

\begin{equation}
    \bar{S}_k^{i^{-1}} \bar{d}_{k-1}^i=\sum_j^{\mathcal{N}_i} \omega_j \hat{S}_k^{j^{-1}}\hat{d}_k^j
\label{eq:input_fusion}
\end{equation}

where weights are decided by faster CI which avoids complex matrices manipulation and non-linear optimization.

\begin{equation}
\omega_j=\frac{1 / \operatorname{tr}{\hat{S}}_j}{\sum_{n=1}^{\mathcal{N}_i} 1 /\operatorname{tr}{\hat{S}}_n}
\end{equation}

\begin{equation}
    \bar{S_k}^i=\left(\sum_{j \in \mathcal{N}_i} w_j \hat{S}_k^{j^{-1}}\right)^{-1}
\label{eq:input_fusion_p}
\end{equation}

Given that all input estimations $\hat{d}^j$ are unbiased, the fusion process described in Eq.(\ref{eq:input_fusion}) remains unbiased. It's important to note that this covariance fusion method does not affect future input variance predictions.

\begin{figure}[htb]        
\centering
\subfloat[Abrupt measurements induce huge bias in input estimates (left) and then influence state (right).] {
   \includegraphics[width=0.95\columnwidth]{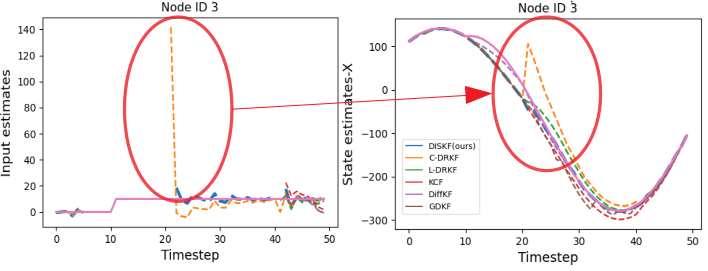}
} \\ 
\subfloat[Reset time window counting during input fusion improves estimations]{
   \includegraphics[width=0.95\columnwidth]{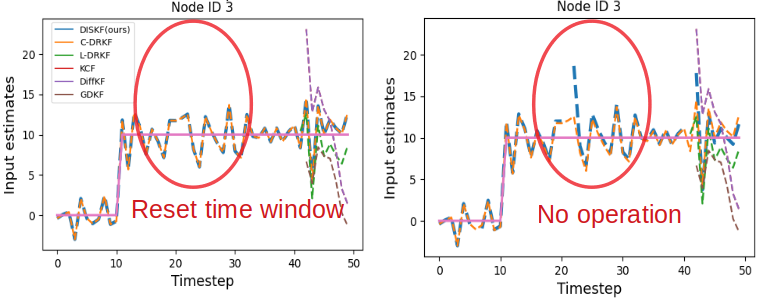}
}
\caption{Observation time window helps deal with intermittent observations in dynamic topology. (a) the sensitivity issue is mitigated by implementing a review through the observation window before estimating inputs. (b) the time window counting should be reset when valid input estimations are available from neighborhoods.}
\label{fig:ttwindow}
\end{figure}

In scenarios when moving agents with dynamic communication topology, the target is intermittently observed. If input estimation is constantly employed, abrupt measurements may be misinterpreted as input influences affecting state estimates, which leads to a significant bias. To mitigate this, we propose a specific operation: establishing an observation time window review mechanism, the length of the window can be regarded as a hyperparameter. If the gap between two consecutive measurements exceeds this window, the initial estimation upon receiving new observations will exclude input estimation. After implementing a review through the observation window before input estimation, our algorithm can mitigate this sensitivity issue. However, the time window should be reset to zero if valid input estimates from neighborhoods are available during the input fusion stage. The success of this approach is demonstrated in Fig. \ref{fig:ttwindow}.

\section{State fusion via information filter decomposition}
Rather than iterative state exchange for consensus among agents, state fusion leverages information Kalman filter decomposition\cite{durrant1990toward}. This fusion approach ensures performance equivalent to an agent with complete neighborhood information in both state estimation and covariance matrix.

After input fusion, yielding fused input estimation $\bar{d}^i_{k}$ and covariance $\bar{S}_k^i$, we proceed from an information Kalman filter perspective, incorporating all node data within $\mathcal{N}_i$. The final state estimation relies solely on $\bar{d}^i_k$, $\hat{x}^i_{k \mid k-1} $, $\hat{P}^i_{k \mid k-1}$, $\tilde{x}^i_{k \mid k}$, $\tilde{P}_{k \mid k}^{i}$.

In order to avoid symbol chaos of equations, we use superscript $*$ on states and covariance to indicate the state fusion process in contrast to the process before communication. For node $i$, incorporating input fusion into the prediction process yields:
\begin{equation}
\begin{aligned}
\hat{x}^{*^i}_{k \mid k-1} & =A^i_{k-1} \bar{x}^{i}_{k-1 \mid k-1}   + G^i_{k-1} \bar{d}^i_{k-1}\\
\hat{P}^{*^i}_{k \mid k-1}
& =A^i_{k-1}  \bar{P}^{i}_{k-1 \mid k-1} A^{i^T}_{k-1}+Q^i_k
\end{aligned}
\label{eq:info_kf_prediction}
\end{equation}

Here, the covariance update is the same as Eq. (\ref{eq:info_kf_prediction0}). This operation is justified because the input covariance impacts only the Kalman filter's gain matrix, potentially leading to non-unique Kalman gains and additional computational load. Previous research\cite{gillijns2007unbiased, kitanidis1987unbiased} indicates that omitting input covariance injection still yields near-optimal Kalman gain. Moreover, exact state-input correlation would require full distributed agent system knowledge, deviating from our objective. Various consensus algorithms\cite{kamgarpour2008convergence,  olfati2008distributed, kamal2012information, ma2016gossip} demonstrate that fusion with approximate covariance can achieve satisfactory performance, hence this paper proceeds without detailed proof of covariance fusion's impact.



Assuming a virtual information filter with $H_k$, $y_k$, and $R_k$ representing aggregated matrices from $\mathcal{N}i$, we simulate all observations being accessible to node $i$. Importantly, we show the final fusion result does not depend on external observations or matrices. Maintaining the prediction from Eq.(\ref{eq:info_kf_prediction}), the update process follows:

\begin{equation}
\begin{aligned}
\bar{x}^{i}_{k \mid k}
& =\hat{x}^{*^i}_{k \mid k-1}+W_{k}\left[y_{k}-H_{k}  \hat{x}^{*^i}_{k \mid k-1}\right] \\
\bar{P}_{k \mid k}^{{i}^{-1}}
& =\hat{P}^{{*^i}^{-1}}_{k \mid k-1}+H_{k} R_{k}^{-1} H_{k} \\
W_{k}
& =\bar{P}^{i}_{k \mid k} H_{k}^{\top}  R_{k}^{-1} \\
\end{aligned}
\label{eq:info_kf_update}
\end{equation}


Substitute Eq.(\ref{eq:info_kf_prediction}) to Eq.(\ref{eq:info_kf_update}),
\begin{equation}
\begin{aligned}
\bar{x}^{i}_{k \mid k}
& = \bar{P}_{k \mid k}^{i} [ H_{k}^{\top} R_{k}^{-1}  y_{k} +  \hat{P}^{{*^i}^{-1}}_{k \mid k-1}\hat{x}^{*^i}_{k \mid k-1}]
\end{aligned}
\end{equation}

After unpacking the observation vector,
\begin{equation}
\begin{aligned}
    H_k^{\top} R_k^{-1} y_k=\sum_{j \in \mathcal{N}} H_k^{j T} R_k^{j^{-1}} y_k^j
\end{aligned}
\end{equation}
\begin{equation}
\begin{aligned}
    H_k^{\top} R_k^{-1} H_k=\sum_{j \in \mathcal{N}} H_k^{j T} R_k^{j^{-1}} H_k^j
\end{aligned}
\end{equation}



Up to now, the state fusion is derived as:
\begin{equation}
\begin{aligned}
\bar{x}_{k \mid k} ^{i}
 = \bar{P}_{k \mid k}^{i} [ \sum_{j \in \mathcal{N}}( \tilde{P}_{k \mid k}^{j^{-1}} \tilde{x}_{k \mid k}^{j} &- \hat{P}_{k \mid  k-1}^{j^{-1}} \hat{x}_{k \mid {k-1   }}^{j} ) \\
&+ \hat{P}^{{*^i}^{-1}}_{k \mid k-1}(\hat{x}_{k \mid k-1}^{i} + G^i_{k-1} \bar{d}^i_{k-1})  ]
\end{aligned}
\label{eq:state_fusion}
\end{equation}

\begin{equation}
\begin{aligned}
\bar{P}_{k \mid k}^{i^{-1}}
& =  \hat{P}^{{*^i}^{-1}}_{k \mid k-1}+
\sum_{j \in \mathcal{N}} (\tilde{P}_{k \mid k}^{j^{-1}} - \hat{P}_{k \mid {k-1}}^{j^{-1}})
\end{aligned}
\label{eq:state_fusion_p}
\end{equation}

Initially, agents estimate $\hat{d}^i_k$, $\hat{S}^i_k$, $\hat{x}^i_{k \mid k-1} $, $\hat{P}^i_{k \mid k-1}$, $\tilde{x}^i_{k \mid k}$, $\tilde{P}_{k \mid k}^{i}$ independently. Post-communication, input and state fusion follows Eq.(\ref{eq:input_fusion}), Eq.(\ref{eq:input_fusion_p}), Eq.(\ref{eq:state_fusion}), and Eq.(\ref{eq:state_fusion_p}), outlined in Algo.\ref{algorithm1}. As this decomposition shows the exact same as the central estimator, the final fused states also have an unbiased and minimum-variance guarantee. 

For non-all-to-all and dynamic communication topologies, a diffusion operation post-Eq.(\ref{eq:state_fusion}) helps correct state estimation errors:

\begin{equation}
    \bar{x}_{k \mid k} ^{i} = \bar{x}_{k \mid k} ^{i} + \epsilon M_i \sum_{j \in \mathcal{N}} ( \hat{x}^{*^j}_{k \mid k-1} - \hat{x}^{*^i}_{k \mid k-1})
\label{eq:compensation_state}
\end{equation}

with $M_i =\bar{P}_{k \mid k}^{i}$ and $\epsilon$ set to 0.05. This state adjustment, averaging predictions, differs from consensus or diffusion KF, which requires further information exchange post-fusion. Our decomposed algorithm naturally zeros out such an adjustment in all-to-all communication networks, offering advantages over consensus-based filters. The ablation study of the state compensation is presented in Table \ref{tab:state_compensation}.

\begin{algorithm}[htb]  
\caption{Decentralized input and state estimation on node $i$ based on information filter }
\label{algorithm1}
\SetAlgoLined
\SetKwData{Left}{left}\SetKwData{This}{this}\SetKwData{Up}{up}
\SetKwFunction{Union}{Union}\SetKwFunction{FindCompress}{FindCompress}
\SetKwInOut{Input}{input}\SetKwInOut{Output}{output}

\Input{Given system models $A^i_k$, $G^i_k$, $H^i_k$,$Q^i_k$,$R^i_k$, dynamic 1-hop neighborhoods $\mathcal{N}^i$.
}
\Output{The input fusion $\bar{d^i_k}$, state fusion $\bar{x}_{k \mid k} ^{i}$, and their respective covariance $\bar{S_k}^i$ and $\bar{P}_{k \mid k}^{i}$.}

Initialize $ x_0^i  = x_0$.
\\ 
\For{timestep $k=1,2,...$}{
1.  \textbf{Individual estimation}

All agents independently perform state estimation as Eq.(\ref{eq:info_kf_update0}). Check if the measurement gap is less than a time window and perform input estimation as Eq.(\ref{eq:input_estimation}).

2. \textbf{Communication}

Exchange and collect information with available nodes within neighborhoods $\mathcal{N}^i$.\\
$\bar{d}^i_k$, $\bar{S}^i_k$, $\hat{x}^i_{k \mid k-1} $, $\hat{P}^i_{k \mid k-1}$, $\tilde{x}^i_{k \mid k}$, $\tilde{P}_{k \mid k}^{i}$

3. \textbf{Input fusion} \\
Perform input fusion as Eq.(\ref{eq:input_fusion}) and Eq.(\ref{eq:input_fusion_p}). \\
      
4. \textbf{State fusion} \\
Perform state fusion as Eq.(\ref{eq:state_fusion}) and Eq.(\ref{eq:state_fusion_p}),\\
add state compensation as Eq.(\ref{eq:compensation_state}).
%


}
\end{algorithm}

\section{Simulation and Experiment Result}
\label{algo:1}
\subsection{Experiment Setup}
\begin{figure}[htb]        
 \center{\includegraphics[width=8.4cm]  {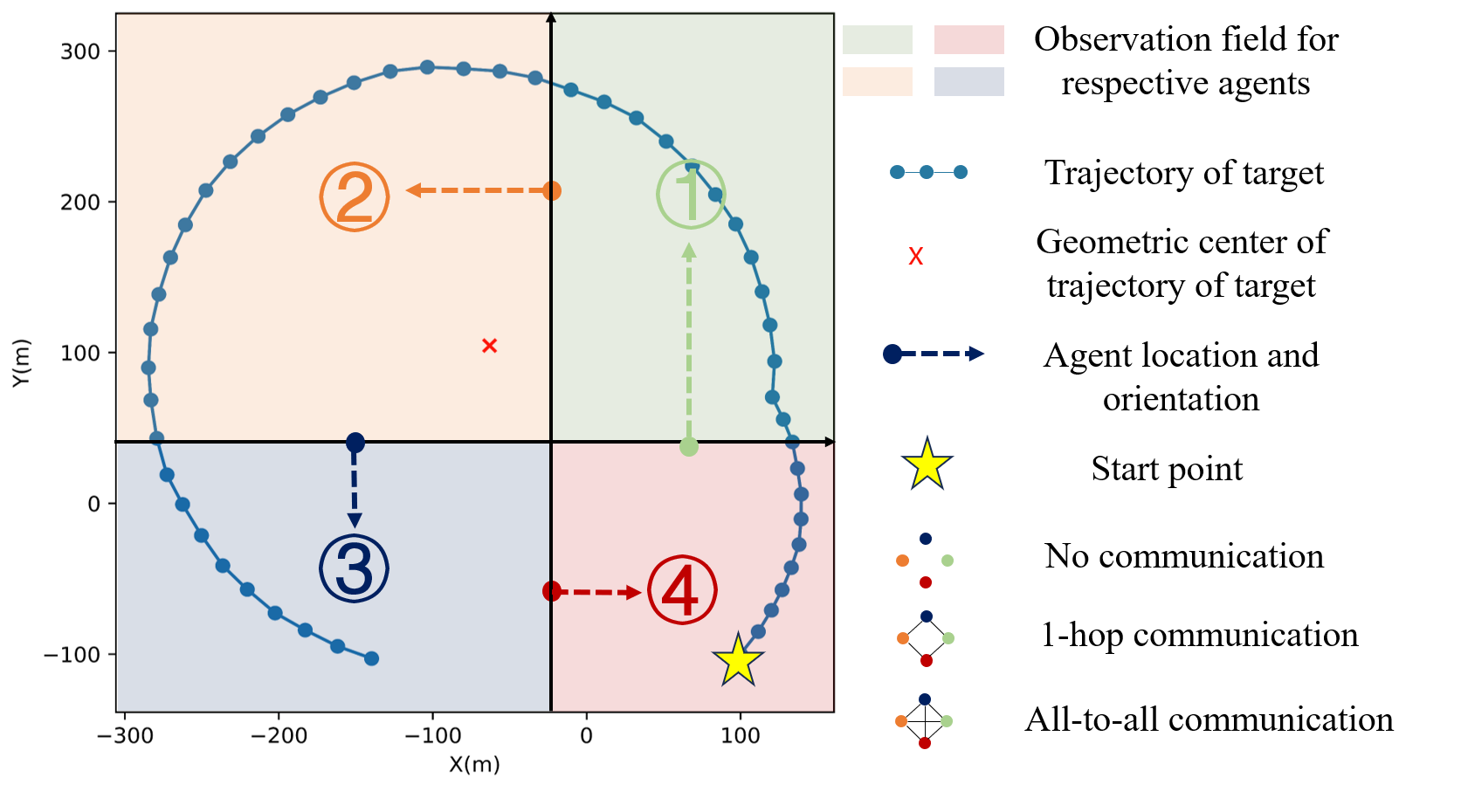}}     
 \caption{\label{demo_1}
Scenario 1: Four static agents are located in the four quadrants. Two of them can observe only the y-coordinate of the target, while another two are limited to observing only the x-coordinate. Each agent is tasked with observing the target exclusively when it enters its respective quadrant. The communication is established according to three distinct typologies.}
\setlength{\belowcaptionskip}{-1cm}  
\vspace{-0.2cm} 
\end{figure}

\begin{figure}[htb]        
 \center{\includegraphics[width=7.4cm]  {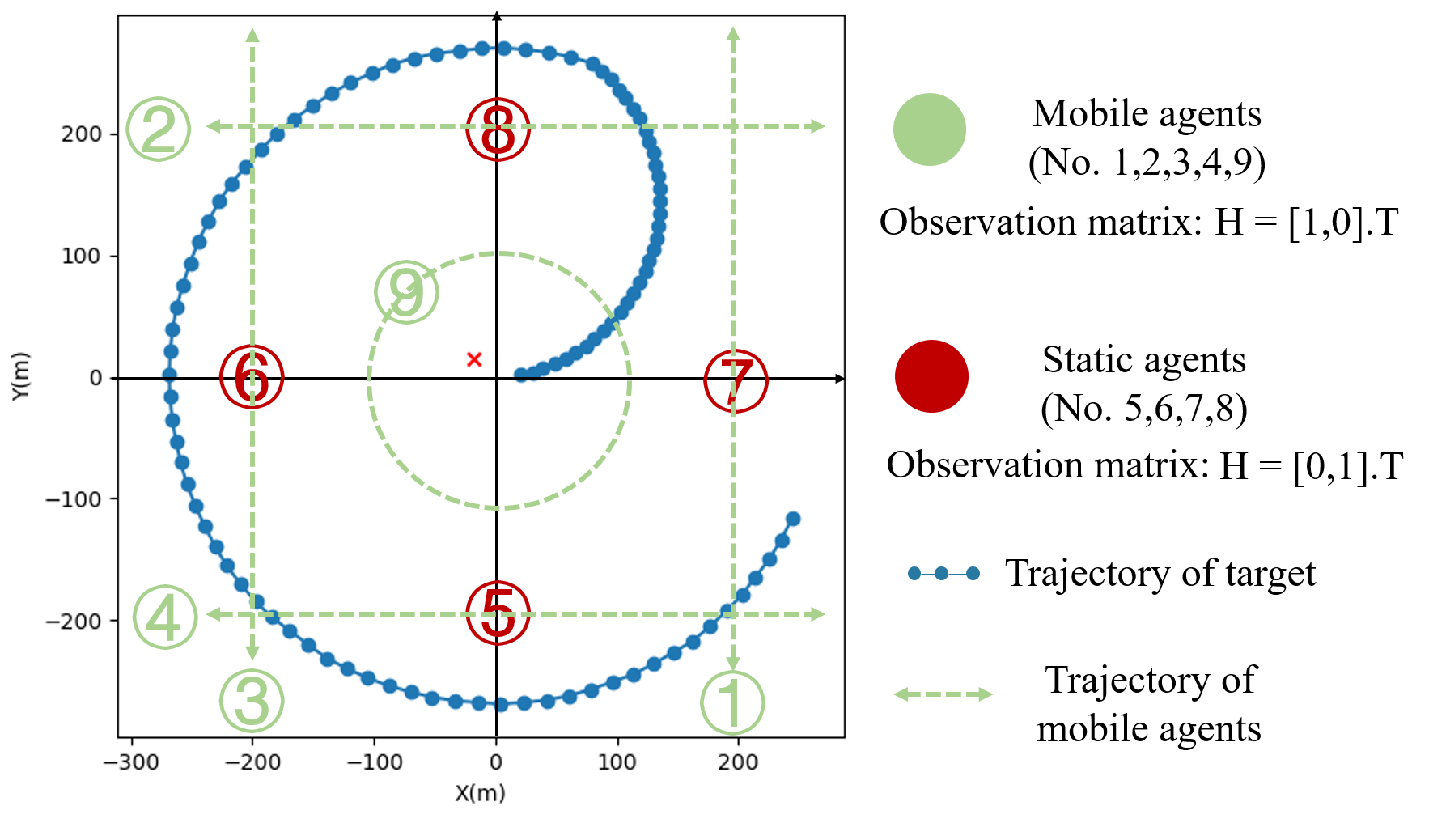}}     
 \caption{\label{demo_2}
Scenario 2: 4 static (red) agents and 5 mobile (green) agents compose a dynamic system. Observation of the target by an agent is contingent on sharing the same quadrant with the target, with each agent limited to observing a single target dimension. The trajectories of the mobile agents are delineated by green dashed lines. Each agent possesses a pre-defined communication range, enabling information exchange solely when the communication ranges of two agents intersect.}
\setlength{\belowcaptionskip}{-1cm}  
\vspace{-0.2cm} 
\end{figure}

In this section, we describe two challenging experiments inspired by a canonical scenario, where a target moves around a base (e.g., the coordinate origin), with its movement influenced by internal unknown inputs, such as sensor faults. Consequently, the trajectory's geometric center deviates from the coordinate origin, as depicted in Fig.\ref{demo_1}. The target's path is segmented into four quadrants, each monitored by a sensor that detects only one dimension of the target's position. All experiment results shown in this paper are carried on under 20 different seeds and averaged. Table \ref{tab:statistics} indicates the average of statistics of the first experiment. Fig. \ref{fig:dynamic_topo} shows the MAE, RMSE as well as variance from all seeds of the second dynamic topologies. Table \ref{tab:state_compensation}
 showcases the average statistics of state compensation in both two experiments.
In the first example, agents are static to observe a moving target. Each of the agents has a different observation model. Specifically, Sensor No.1 and No.3 can measure the $x$ position in its quadrant, $H_1 = H_3 = [1, 0]$, while Sensor No.2 and No.4 will measure the $y$ position, $H_2 = H_4 = [0, 1]$. These sensors ensure at any moment, the target is tracked by only one sensor for a singular dimension. During this experiment, observations from a central agent are absent akin to environmental monitoring practices\cite{ben2023multi}.  

In the second example, the sensor network consists of two groups: dynamic and static sensors. Most settings are similar to the first example. In particular, the communication topology is fully dynamic, as each agent has its own communication field. A communication link is established only when agents are within range to communicate. The communication range in this paper is 120, the detail is seen in Fig. \ref{demo_2}.

The system matrices describe a dynamic model that makes the target move in a circular trajectory. 

\begin{equation}
A_k=\left[\begin{array}{cc}
\cos \left(\theta_k\right) & -\sin \left(\theta_k\right) \\
\sin \left(\theta_k\right) & \cos \left(\theta_k\right)
\end{array}\right] 
\end{equation}
where $\theta_k = \omega  +\theta_0$ ,$\omega$ is the angular velocity. $k$ is the timestep, $\theta_0$ is the initial angle.
\begin{equation}
    \quad G_k=\left[\begin{array}{cc}
1  \\
1 
\end{array}\right]\quad B_k=\left[\begin{array}{l}
d(t)
\end{array}\right]
\end{equation}
where $d(t)$ is the internal unknown input that cannot be predicted and observed directly, like the internal sensor fault in a sensor network. To mimic the changes of inputs over time. The function of $d(t)$ is:
\begin{equation}
    d(t)= \begin{cases}0, & \text { if } k<10 \\ 10, & \text { if } k \geq 10\end{cases}
\end{equation}

\begin{equation}
    Q_i = [0.2, 0.2]^T; R_i = [2]^T; i \in\{1,2,3,4..,N\}
\end{equation}

Our algorithm is denoted as DISKF (Decentralized Input and state Kalman filter), and the baselines are \cite{liu2018minimum} which assume agents can have a global view of system equations of neighborhoods, as it seems like a locally centralized filter, so we call it C-DRKF; L-DRKF \cite{peng2023optimal,emami2020distributed} which do a state consensus but with individual input estimation; KCF\cite{ma2016gossip} which is the representative of consensus algorithms which will take a weighted average of exchanged states; DiffKF\cite{cattivelli2010diffusion} which is the representative of diffusion-based algorithms which will average states of neighborhoods after each agent performs a consensus of states; GDKF \cite{petitti2011consensus} is a gossip-based algorithm which will randomly and asynchronously collect message from neighborhoods. For fairness, all algorithms are only allowed to communicate once between two consecutive timesteps. 

\begin{figure}[htbp]        
 \center{\includegraphics[width=8.4cm]  {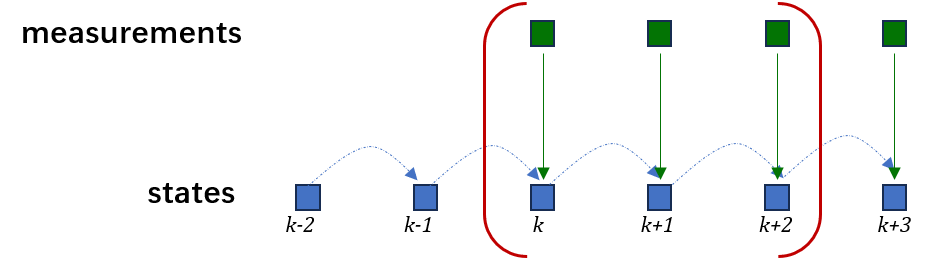}}     
 \caption{\label{inputs}
 The time window review deals with the abrupt bias caused by input estimation facing intermittent measurements. When the sensor abruptly measures the target, within the period of the time window review, only do state estimation without input estimation. 
 }
\setlength{\belowcaptionskip}{-1cm}  
\vspace{-0.2cm} 
\end{figure}

\subsection{Result}
Given the page constraints, we only provide a summary of the 4-agents case with 1-hop communication and the dynamic scenarios with 9 agents. 

Firstly, we illustrate input estimation using 1-hop communication in Fig.\ref{inputs}, whereas some baselines like KCF, DiffKF, and GDKF lack detailed input estimation in their documentation. For fairness, we adopted the RKF's input estimation method\cite{gillijns2007unbiased} for comparison with these algorithms, noting that this adaptation only serves for comparison but didn't affect state estimation; the original algorithms' state estimations were preserved.

In 1-hop limited communication where only one sensor observes the target at any time, no agent can have constant information about the target, thus leading to missing data in Fig.\ref{inputs}. Algorithms without specialized input designs, like KCF, DiffKF, and GDKF, exhibit significant bias However, filters with comprehensive information of neighbors may also face challenges, as indicated by the bias in input estimation for nodes 3 and 4 when the target enters quadrants 3 and 4. This occurs because before entering quadrant 3, node 3's state and covariance evolve fully depending on the prediction model which makes state error significantly increases, whereas node 2's estimation adjusts upon the target's entry into quadrant 1. Based on Eq. (\ref{eq:input_estimation}), the input estimation depends on the observation, current states, and its covariance. Then if 
using node 2's observation but with node 3's state, the estimator will misinterpret the innovation as the influence of unknown input. This is a similar issue we discussed in the observation time window chapter. However, without state and input fusion, even the C-DRKF with full observation and corresponding matrices still cannot solve this issue. Our fusion method, considering both input and state along with their covariances, maintains stable input estimation.

\begin{figure}[htbp]        
 \center{\includegraphics[width=8.4cm]  {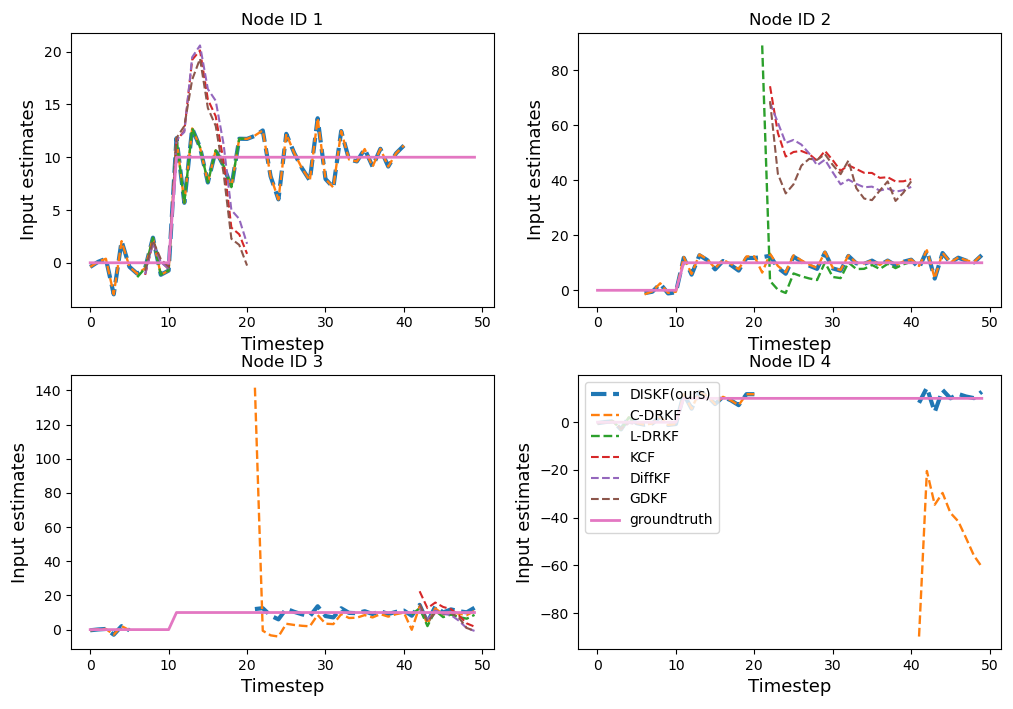}}     
 \caption{\label{inputs}
 The input estimation with 1-hop communication. The index of nodes corresponds to Fig.\ref{demo_1}. Missing estimation means the corresponding algorithm can't recover the input from the data they observe or exchange.
 }
\setlength{\belowcaptionskip}{-1cm}  
\vspace{-0.2cm} 
\end{figure}

\begin{figure}[htb]        
 \center{\includegraphics[width=8.4cm]  {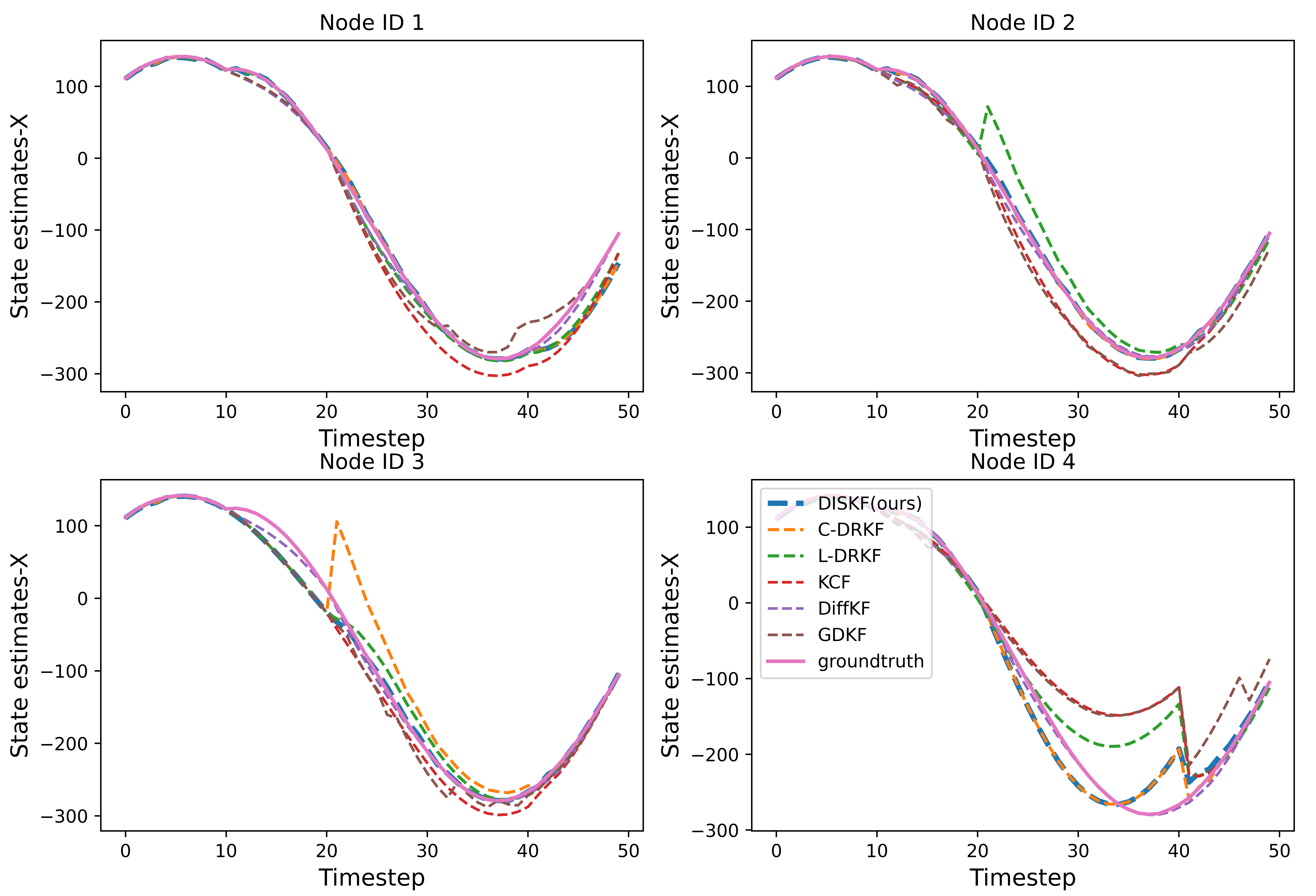}}    
 \caption{\label{states_x}
The $x$ coordinate of position estimation of the target with 1-hop communication. 1) Our algorithm (blue line) can converge back to groundtruth very quickly once one of the nodes within 1-hop has the observation of the target. 2) even with limited information exchange, our algorithm is still very close to filtering (blue line) with full information. 
 }
\setlength{\belowcaptionskip}{-1cm}  
\vspace{-0.2cm} 
\end{figure}

\begin{figure}[htb]        
 \center{\includegraphics[width=8.8cm]  {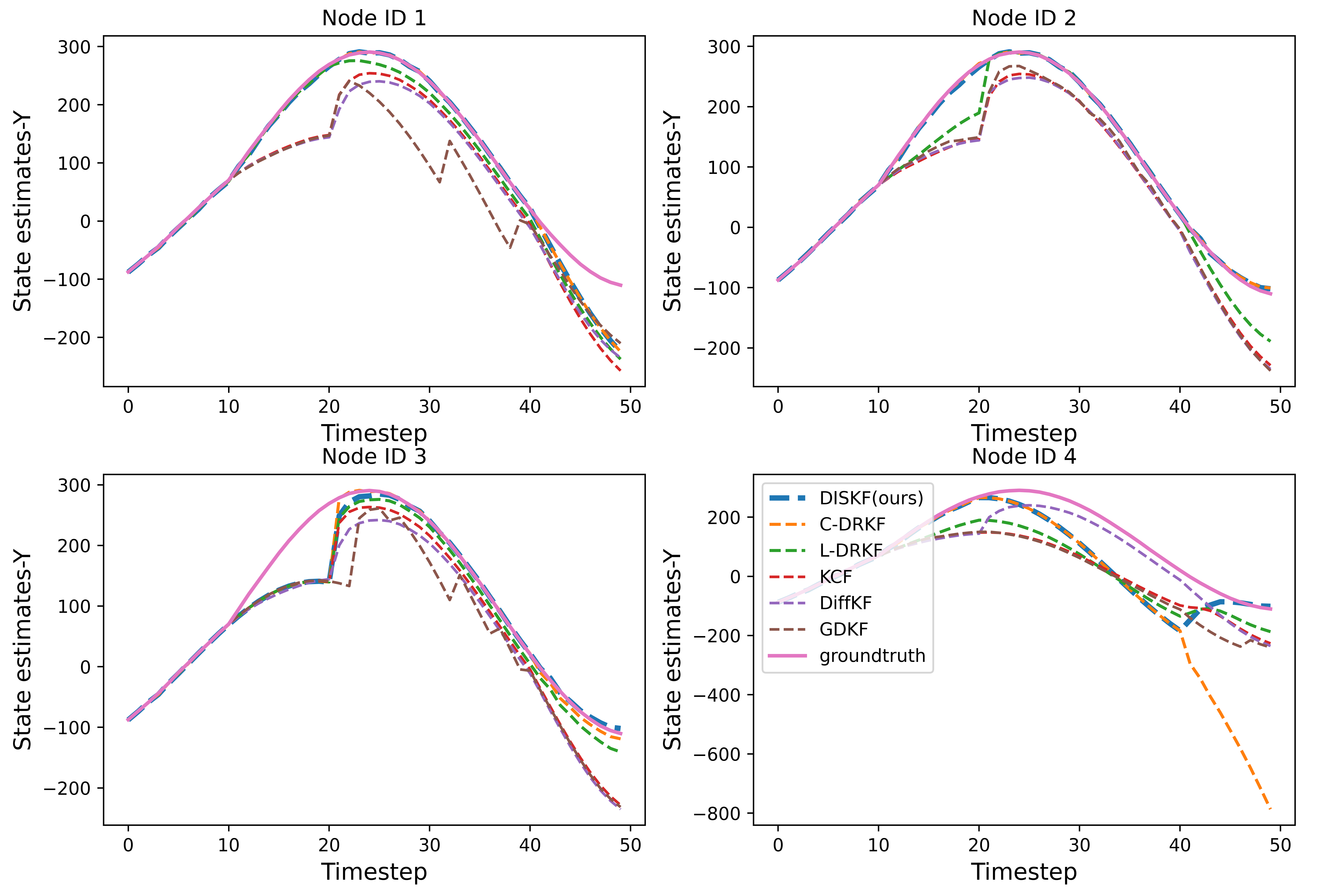}}    
 \caption{\label{states_y}
The $y$ coordinate of position estimation of the target with 1-hop communication.
 }
\setlength{\belowcaptionskip}{-1cm}  
\vspace{-0.2cm} 
\end{figure}

State estimations are depicted in Fig.\ref{states_x} and Fig.\ref{states_y}, demonstrating that a sensor's singular dimensional observation leads to asynchronous $x$ and $y$ estimation trends. Algorithms lacking input estimation invariably present bias in state estimations. While Kalman filters can naturally compensate for sudden disturbances over time, continuous or lasting inputs can still introduce bias during estimation.

\begin{figure}[htb]        
 \center{\includegraphics[width=7.7cm]  {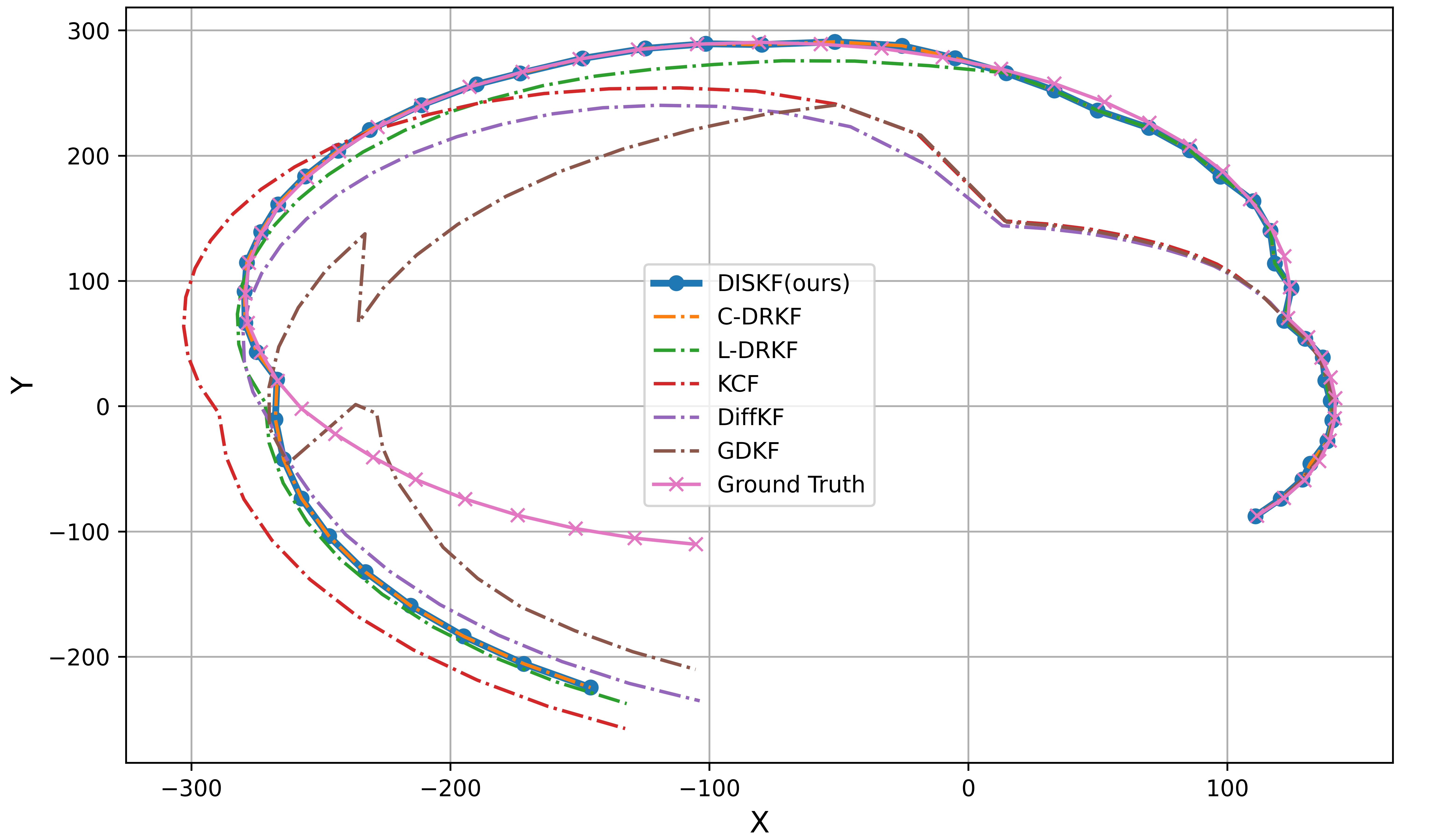}}
 \caption{\label{traj}
2D trajectory visualization of all estimators with 1-hop communication. Our algorithm (blue line) exhibits robustness against unexpected internal input disturbance while maintaining performance parity with the filter that possesses complete neighborhood information.
 }
\setlength{\belowcaptionskip}{-1cm}  
\vspace{-0.2cm} 
\end{figure}

Estimated trajectories in Fig.\ref{traj} showcase our algorithm's robustness in scenarios marked by incomplete observations and unknown internal inputs. Despite limited information, our algorithm compares favorably or even outperforms others.

\begin{figure}[H]        
 \center{\includegraphics[width=8.4cm]  {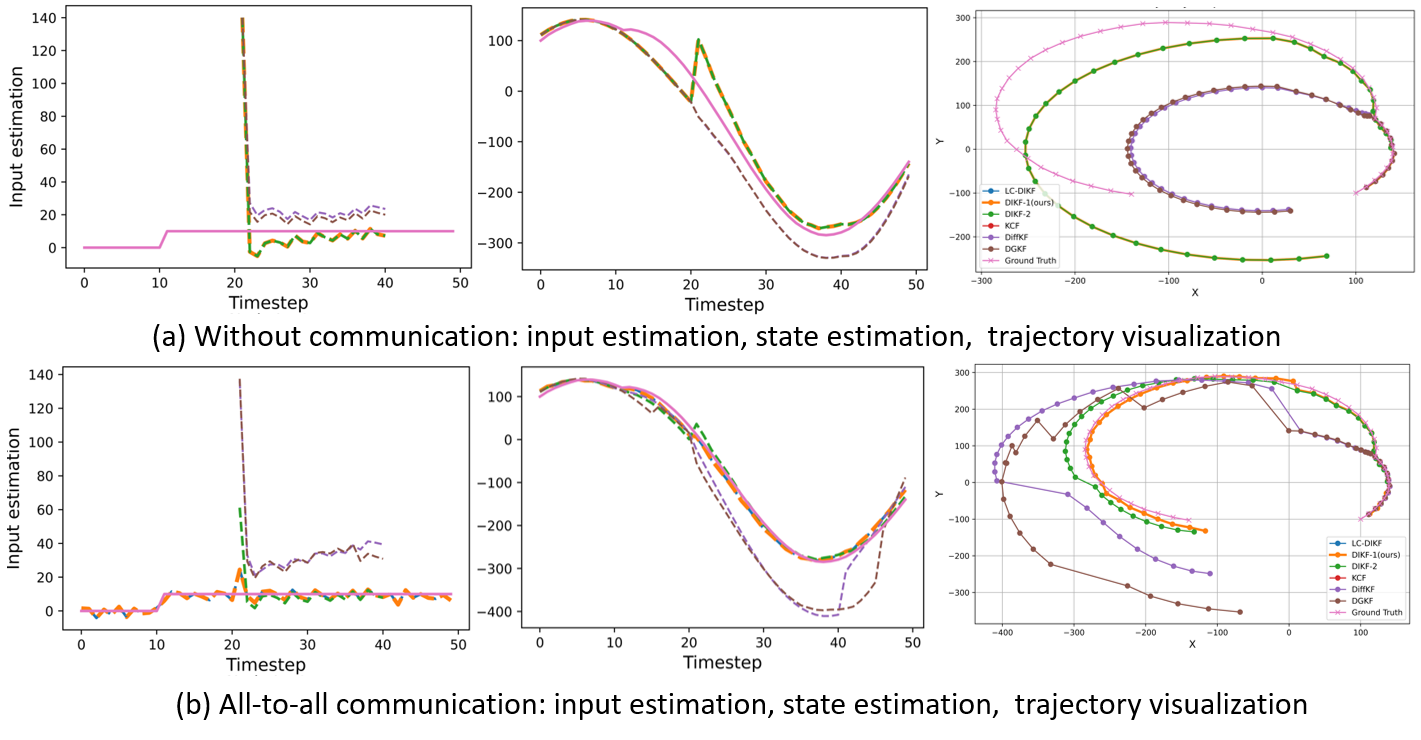}}    
 \caption{\label{sub}
Supplementary results of communication without communication and all-to-all communication of node 2.}
\setlength{\belowcaptionskip}{-1cm}  
\vspace{-0.2cm} 
\end{figure}

Additionally, we examine input and state estimations across three different topologies in Fig.\ref{demo_1}, using agent node 2 as an example. Fig.\ref{sub} illustrates that isolated estimations by each agent yield comparable performance among algorithms with and without input estimation. However, under all-to-all communication, our algorithm (blue line) mirrors the performance of the fully informed C-DRKF filter, aligning with our theoretical expectations, which can also shown in Table \ref{tab:statistics}.

\begin{figure}[htb]
\centering
\subfloat[Input estimation (dynamic)] {
   \includegraphics[width=0.45\columnwidth]{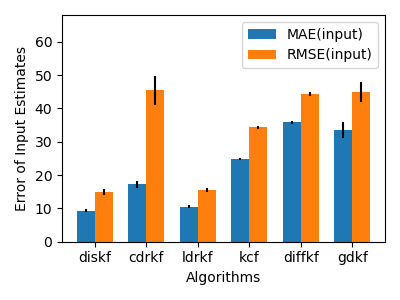}
}
\subfloat[State estimation  (dynamic)]{
   \includegraphics[width=0.45\columnwidth]{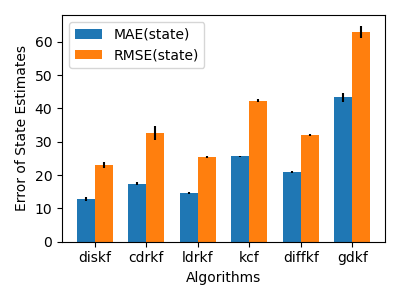}
}\\ 
\subfloat[Input estimation (all-to-all)\,]{
   \includegraphics[width=0.45\columnwidth]{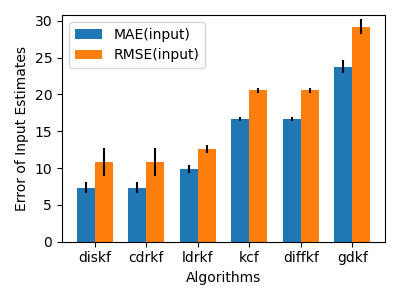}
}\,\,
\subfloat[State estimation (all-to-all)]{
   \includegraphics[width=0.45\columnwidth]{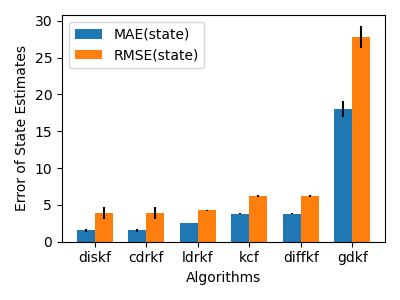}
}
\caption{DISKF(ours) shows superiority over all baselines in multi-agent case under dynamic topology with heterogeneous sensors. (a)(b): Input and state estimation error with dynamic communication topology (c): Input and state estimation error with all-to-all communication}
\label{fig:dynamic_topo}
\end{figure}

\begin{table}[H]
\centering
\caption{Evaluation of State Compensation Effect}
\label{tab:state_compensation}
\resizebox{8.4cm}{11.8mm}{
\small 
\begin{tabular}{|c|c|c|c|c|c|c|}
\hline
Env. & Comm. & State Comp. & MAE (input) & RMSE(input) & MAE(state)& RMSE(state) \\
\hline
& \multirow{2}{*}{All-to-all} & Yes  & 4.63 & 6.35  &1.91  & 2.66   \\
& & No & 4.63 & 6.35  &1.91  & 2.66   \\
\cline{2-7}
Stationary & \multirow{2}{*}{1-hop} & Yes  & \textbf{15.76} & \textbf{30.05}  & \textbf{4.67}  &\textbf{6.37 }   \\
& & No & 26.75  & 55.63 &7.38  & 15.36  \\
\hline
& \multirow{2}{*}{All-to-all} & Yes  & 1.56 & 4.01  &7.40  & 11.08   \\
& & No  & 1.56 & 4.01  &7.40  & 11.08   \\
\cline{2-7}
Dynamic & \multirow{2}{*}{1-hop} & Yes  & \textbf{10.37} & \textbf{19.34}  & \textbf{8.82}  &\textbf{14.22}   \\
& & No & 16.14  & 26.47 &9.11  & 14.76  \\
\hline
\end{tabular}}
\end{table}

\begin{figure}[H]        
\centering
\subfloat[Input estimation versus radius] {
   \includegraphics[width=0.47\columnwidth]{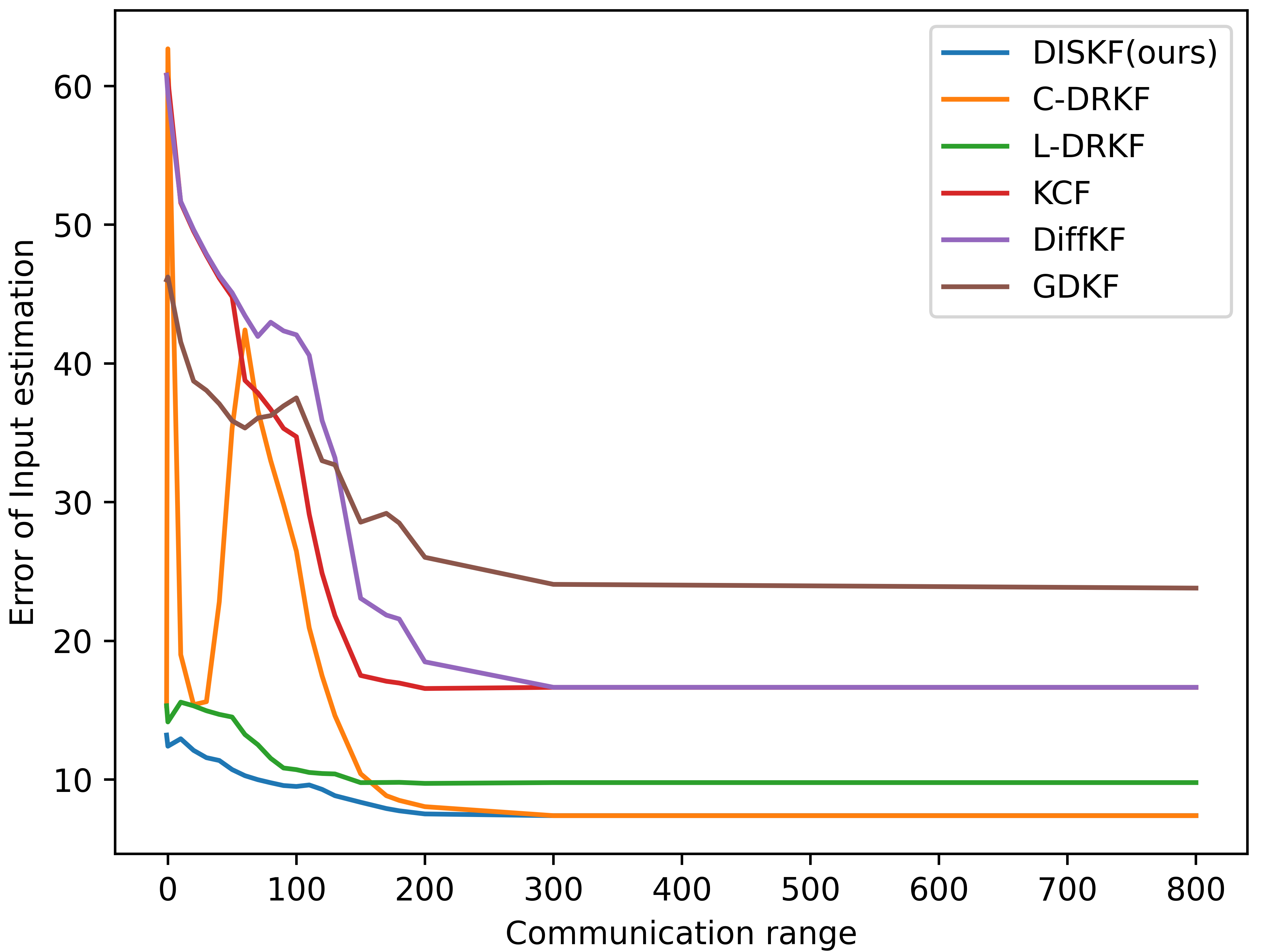}
} 
\subfloat[State estimation versus radius]{
   \includegraphics[width=0.48\columnwidth]{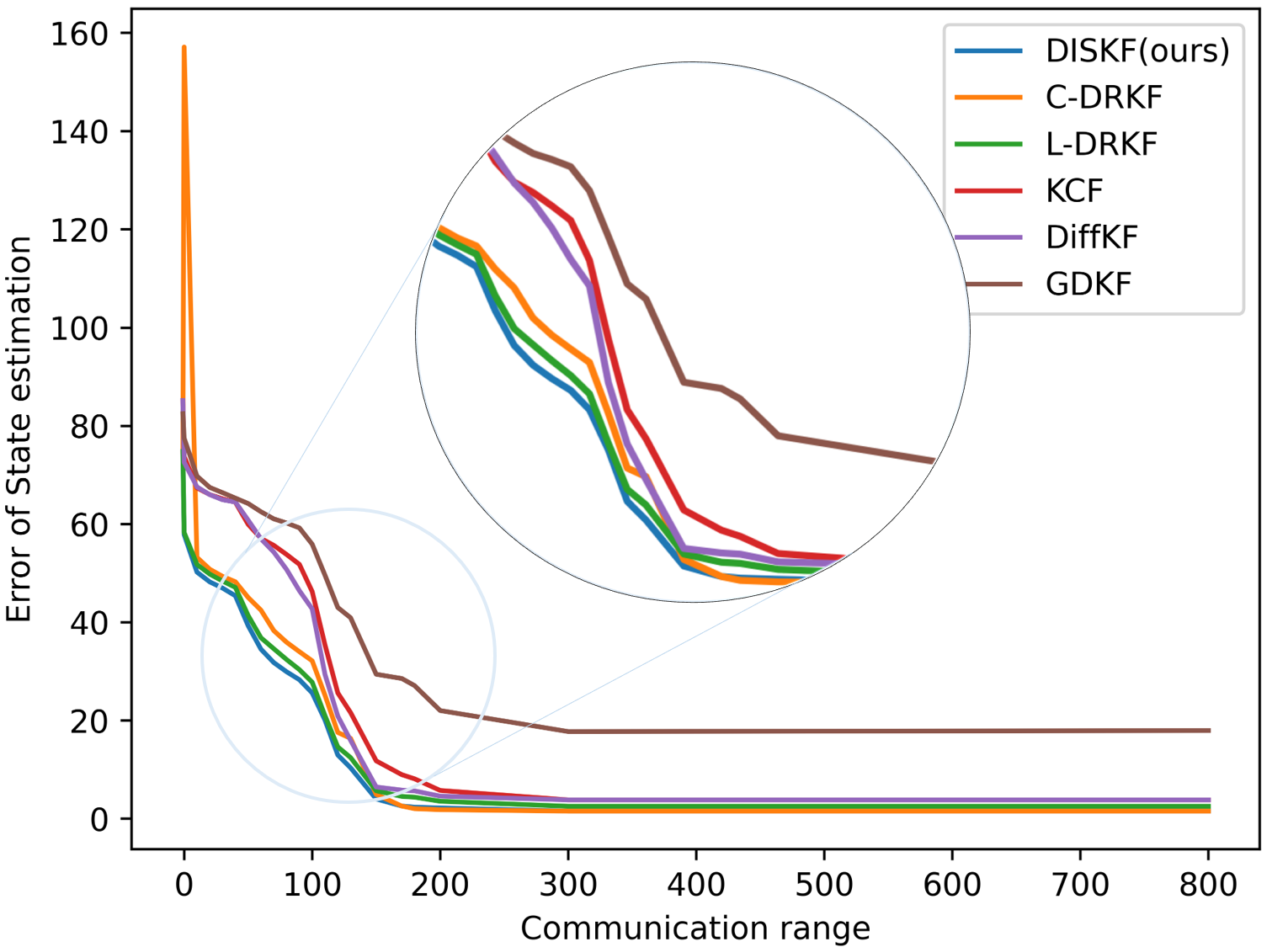}
}
\caption{Algorithm performance versus communication ranges in the dynamic topology scenario. (a) and (b) show the Mean Absolute Error (MAE) of input and state estimation under different communication ranges of each agent.  
}
\label{fig:error_vs_radius}
\end{figure}

Table \ref{tab:statistics} displays statistical data for scenarios involving 4 stationary agents, evaluated by Mean Absolute Error (MAE) and Root Mean Squared Error (RMSE) metrics. The results demonstrate enhanced estimation accuracy with increased communication, with our algorithms surpassing all baseline models consistently. Figure \ref{fig:dynamic_topo} showcases statistical outcomes for a dynamic system, highlighting dynamic topology with a 120 communication range and all-to-all communication. This figure reveals that our algorithm not only excels in performance within the dynamic topology but also matches the performance of C-DRKF in scenarios of all-to-all communication. The C-DRKF, requiring a global perspective of other agents, represents global optimality in all-to-all communication settings. Our algorithm, requiring only suitable information exchange yet achieving comparable performance, is remarkable. In addition, an in-depth analysis explores the impact of communication range, as illustrated in Fig. \ref{fig:error_vs_radius}. This figure confirms that our algorithm consistently outperforms across different communication ranges. Unlike the C-DRKF algorithm, which experiences unstable estimation due to intermittent observation issues in dynamic topologies, our algorithm effectively manages this issue, aligning with theoretical predictions.

Table \ref{tab:state_compensation} demonstrates the effectiveness of state compensation, as detailed in Eq.(\ref{eq:compensation_state}). As noted earlier, in the scenario where there is all-to-all communication, state compensation naturally reduces to zero—a unique attribute not seen in other consensus algorithms, as the statistical data underlines. In a dynamic communication topology, state compensation improves the accuracy of estimations. The relative advantage of state compensation is also revealed in the analysis between C-DRKF and L-DRKF. Unlike C-DRKF, L-DRKF incorporates an additional operation of state compensation (consensus) beyond input estimation. Hence, despite only relying on local information for input estimation, L-DRKF outperforms C-DRKF in certain situations.

\begin{table}[htb]
\centering
\caption{State and Input Estimation Error in Four Agents Case under Stationary Topology with Heterogeneous Sensors}
\label{tab:statistics}
\resizebox{8.4cm}{15.5mm}{
\small 
\begin{tabular}{|c|c|c|c|c|c|c|c|}
\hline
Comm. & Metric & \textbf{\bfseries DISKF (ours)} & C-DRKF \cite{liu2018minimum} & L-DRKF \cite{peng2023optimal} & KCF\cite{olfati2008distributed} & DiffKF \cite{cattivelli2010diffusion} & GDKF\cite{petitti2011consensus} \\
\hline
\multirow{4}{*}{all-to-all} & MAE(state) & \textbf{\bfseries 1.91} & \textbf{\bfseries 1.91} & 15.66  & 27.90  & 27.90  & 36.20  \\
& RMSE(state) & \textbf{\bfseries 2.66} & \textbf{\bfseries 2.66}  & 25.05  & 40.45  & 40.45  & 50.15  \\
& MAE(input) & \textbf{\bfseries 4.63} & \textbf{\bfseries 4.63} & 6.87  & 15.39  & 15.39  & 14.01 \\
& RMSE(input) & \textbf{\bfseries 6.35} & \textbf{\bfseries 6.35 }  & 9.75  & 16.51  & 16.51  & 15.12 \\
\hline
\multirow{4}{*}{1-hop} & MAE(state) & \textbf{\bfseries 15.76} & 26.76 & 26.08  & 37.40  & 24.87  & 40.92  \\
& RMSE(state) & \textbf{\bfseries 30.05} & 55.63  & 41.54  & 53.41  & 40.84  &58.89 \\
& MAE(input) & \textbf{\bfseries 4.67} & 9.73 & 6.30  & 15.11 & 14.53 & 13.35\\
& RMSE(input) & \textbf{\bfseries 6.37}  & 17.34 & 9.72  & 16.94 & 15.47  & 14.33 \\
\hline
\multirow{4}{*}{None} & MAE(state) & \textbf{\bfseries 63.75}  & 71.01 & 71.01  & 66.43  & 66.43  & 65.82 \\
& RMSE(state) & 92.88 & 109.31 &109.31  & 90.78  & 90.78  & \textbf{\bfseries 90.36} \\
& MAE(input) & \textbf{\bfseries 9.10} & 21.76 & 21.76  & 16.32  & 16.32  & 12.65 \\
& RMSE(input) & \textbf{\bfseries 12.21} & 32.33 & 32.33  & 17.95  & 17.95  & 14.50 \\
\hline
\end{tabular}}
\end{table}

\section{Conclusion}
This paper introduces a decentralized, resilient, and efficient input and state estimation algorithm. Through system decomposition, state compensation, and observation time window review techniques, our algorithm can deal with dynamic multi-agent systems that consist of heterogeneous sensors with intermittent observations for each agent. This enables us to extend to cooperative tracking, ad-hoc sensor networks, and large spatial-temporal systems like geophysical monitoring, with real-time and privacy demand.


\addtolength{\textheight}{-12cm}   


\bibliographystyle{ieeetr}
\bibliography{ref}


\end{document}